\documentclass[a4paper,11pt]{article}   
\usepackage{lineno,hyperref}
\usepackage[utf8]{inputenc}
\usepackage[english]{babel}
\usepackage{amssymb}
\usepackage{ifthen}

\newtheorem{theorem}{Theorem}
\newtheorem{pro}{Proposition}
\newtheorem{corol}{Corollary}

\date{}

\begin{document}

\title{Partial covering arrays for data hiding and
quantization\thanks{The work was funded by Presidium of the Russian
Academy of Sciences (Grant 0314-2015-0011).}}

\author{ Vladimir N. Potapov}

\maketitle

email: vpotapov@math.nsc.ru; 

\begin{abstract}
We consider the problem of finding a set (partial covering array)
$S$ of vertices  of the Boolean $n$-cube having cardinality
$2^{n-k}$ and intersecting with maximum
 number of $k$-dimensional faces.  We prove
that the ratio between the numbers of the $k$-faces containing
elements of $S$ to $k$-faces is less than
$1-\frac{1+o(1)}{\sqrt{2\pi k}}$ as $n\rightarrow\infty$ for
sufficiently large $k$.
  The solution of the problem in the class of linear
  codes is found. Connections between
this problem, cryptography and an efficiency of quantization are
discussed.


\end{abstract}

\section{Introduction}

Let $F_2^n$ be  the set of binary vectors of  length $n$
(hypercube). We consider some problems relating to information
transmission. The first problem is the message transmission over
wiretap channel  \cite{OW}. Consider the following situation.  An
adversary can intercept $n-k$ bits (in random positions) of an
$n$-bit message. The encoder is to be designed to minimize the
adversary's information about the initial data.
 A general
approach for solving this problem is to split hypercube into
$C_1,\dots, C_{2^k}$ sets, for example, by syndromes of some linear
code, and to encode $k$-bit data $x$ by a random $n$-bit word from
$C_x$. Let $\Gamma$ be $k$-dimensional face ($k$-face) defined by
intercepted $n-k$ bits.  Then the adversary is forced to choose
between all $x$ such that $C_x\cap \Gamma\neq \varnothing$. The
encoder needs a partition such that each $k$-face of the hypercube
intersects with as many as possible elements of partition.  In other
words each $C_x$ must intersect as many $k$-faces as possible.

One of a well-known  stegoscheme is based on coding theory. Encoder
changes one or more bits of the initial message in order to the
resulting word has a special syndrome. This syndrome is a hiding
message. It is assumed that the changes of initial message are not
perceptible. However,  data obtained  by    modern methods of coding
images contain control bits of different kind that cannot be
changed. So called wet paper stegoscheme divides  the coordinates
into wet coordinates that can be used for hiding information and dry
coordinates that cannot be altered \cite{Mun}. However alternation
of different wet coordinates corresponds to hugely different
effects. Places of the least significant bits depend of image.
Consider this issue in detail.

The second problem is an efficient quantization of real data. It is
the important stage for lossy compression of images or speechs. From
the nature of the things a part of data values is on the edges of
the quantization intervals. The last bit of such value is the least
significant one for the quality of quantization. Thus useful
stegoscheme should provide possibility of alternation of different
bits for embedding the same message. We conclude that the set of
words generated by stegoscheme must intersect as many $k$-faces as
possible where $k$ is the maximum number of alternating bits.
 A special method
for choosing this least significant bits is used for data hiding in
image and video \cite{WYL}.

It is possible to utilize this redundancy  for data compression
\cite{FRP}.
 Consider an $n$-tuple consisted of the last
bits of  quantized values. Suppose that  each $n$-tuple contains
$k'$ negligible bits. Let $C$ be  some code (array of codewords)
with cardinality $2^{n-k'}$ and let for each $n$-tuple there exists
a codeword such that this $n$-tuple and the codeword  differ only in
negligible bits.
  We will transmit the codeword (rather its number) instead of the initial
$n$-tuple. So, we will truncate $k'$ bits of the $n$-bit message.

Consider a mathematical formulation of the problem. We would like to
construct a minimal code such that for any binary $n$-tuple $v$ and
for each set of $k$ positions (that contains the least significant
bits) there exists a codeword $u$ such that $v$ and $u$ can differ
from each other only in these $k$ positions. I.e., the $k$-face
defined by these positions contains a codeword. We need either to
find such code (covering array) with the least cardinality or to fix
a cardinality of the code and to maximize the number of the fit
positions. Note that this problem is different from a construction
of covering codes. A code $C$ is called $k$-covering code if for
every $n$-tuple there exists a codeword which is different in
certain $k$ positions. But we going to maximize the number of the
fit  $k$-sets.

A perfect solution for the problems  written above would be a set of
codewords containing only one element of each $k$-face of $F^n_2$.
Such sets have cardinalities $2^{n-k}$ and  are called MDS codes.
For the $q$-ary hypercubes ($q$ is a prime power), there are MDS
codes with several code distances. If the cardinality of an MDS code
is $q^{n-k}$ then the code distance is equal to $k+1$. But in the
Boolean hypercube, there exist only two nonequivalent MDS codes: the
parity check code ($k=1$) and  the pair of antipodal vectors
($k=n-1$), for example, $\overline{0}$ and $\overline{1}$ \cite{MS}.
Consequently, we need to find  approximate solutions.

A subset $T$ of the hypercube is called a binary covering array
$\mathrm{CA}(|T|,n,n-k)$ with strength $n-k$  if for each $v\in
F_2^n$ and for any $k$ positions there exists $u\in T$ such that $v$
and $u$ can differ only in these $k$ positions. A survey of
constructions and bounds for cardinalities of covering arrays can be
found in \cite{LKLK}. At this moment, exact bounds are obtained only
for small $n$ or for $k=1,2,n-2,n-1$ and an arbitrary $n$. If
$n>k+1$,
 the cardinality of minimum covering array
 $F(n,k)$ is greater than $2^{n-k-2}(2+\log
 (k+2))$  and $F(n,k)\leq 2^n/(k+1)$. Moreover, it is known that
$F(n,k)\asymp 2^{n-k}\log n$ as $n\rightarrow\infty$ and $n-k$ is
fixed. With the exception of the parity check code mentioned above,
linear codes are not  useful as binary covering arrays because any
other proper linear code does not intersect   with some
$\lceil\frac{n}{3}\rceil$-faces.

We will consider a bit different mathematical problem: to construct
a partial covering array $S_k\subset F_2^n$, $|S_k|=2^{n-k}$ with
the following property. The number of $k$-faces containing elements
of $S_k$ is as large as possible.  In practice  it is convenient to
encode elements of sets of such cardinality. Apparently for the
first time  partial covering arrays or "covering array with budget
constraints" is considered in \cite{HR}. In \cite{Sar} an existence
of  partial covering arrays with some parameters is established by
probabilistic methods.
 Denote by $ \nu_k(S_k)$ the
ratio between the number of $k$-faces that contain elements of $S_k$
and the number ${{n}\choose{k}}2^{n-k}$ of all $k$-faces. In Section
2, we  prove that $ \max \nu_k(S_k)\approx
1-\frac{1+o(1)}{\sqrt{2\pi k}}$ as $n\rightarrow\infty$ for
sufficiently large $k$.

For application in information transmission a device or algorithmic
function that performs a quantization  must be cost-effective. For
example, linear codes  are easily implemented. In Section 3, we
  find precise value of $ \max
\nu_k(S_k)$ for linear sets as $n=2^{k}-1$ and construct
corresponding $S_k$.

Another approach to study suitability of linear codes for similar
tasks was developed in \cite{ZZ}.

In the beginning, we consider a random subset $T$ of the hypercube
with cardinality $2^{n-k}$. We suppose that the elements of $T$ are
selected independently. Since the probability of any $k$-face
$\Gamma$ equals $1/2^{n-k}$, the probability ${\rm Pr}(T\cap
\Gamma=\varnothing)$ equals $(1-\frac{1}{2^{n-k}})^{2^{n-k}}$. Since
$(1-\frac{1}{2^{n-k}})^{2^{n-k}} \rightarrow 1/e$ as
$n\rightarrow\infty$, the following proposition is true.

\begin{pro}\label{pvid10}
  $\lim\limits_{n\rightarrow\infty}{\rm E}\nu_k(T)=1-1/e$,
  where $T\subset E^n$ is a random set, $|T|=2^{n-k}$, and $k$ is fixed.
\end{pro}

\section{Upper bound for partial covering arrays}

In this section we will use definitions and methods from \cite{MS}.
Consider the vector space $\mathbb{V}$ of real-valued functions on
$F_2^n$ with the scalar product
$(f,g)=\frac{1}{2^n}\sum\limits_{x\in F_2^n}f(x){g(x)}$. For every
$z\in F_2^n$ define a {\it character} $\phi_z(x)=(-1)^{\langle x,
z\rangle}$, where  ${\langle x, z\rangle}= x_1z_1+\dots+x_nz_n$.
Here all arithmetic operations are performed on   real numbers. As
is generally known, the characters of the group  form an orthonormal
basis of $\mathbb{V}$.

Let $M$ be the adjacency matrix of the hypercube $F_2^n$. This means
that $Mf(x)=\sum\limits_{y:d(x,y)=1}f(y)$, where $d(x,y)$ is the
Hamming distance. It is well known that the characters are
eigenvectors of $M$. Indeed, we have
$$M\phi_z(x)=  (n-2wt(z))\phi_z(x), $$ where $wt(z)$
is the number of nonzero coordinates of $z$.

Let $M_r$ be the adjacency matrix of the distance $r$, i.e.
$M_r(x,y)=1$ iff $d(x,y)=r$. Obviously, this matrix generates Bose
--- Mesner algebra. We have the equations
$$M_rM=(n-r+1)M_{r-1}+(r+1)M_{r+1},$$
$$M^r=a^r_0M_0+a^r_1M_1+\dots+a^r_rM_r,$$
where
\begin{equation}\label{evid1}
a^{r+1}_i=ia_{i-1}^r+(n-i)a^r_{i+1},
\end{equation} $a_{-1}^r=a_{r+1}^r=0$.
The following properties of coefficients $a^r_i$
 are not difficult to prove by induction using (\ref{evid1}).

1. $a^r_r=r!$.

2. $a^r_i=0$ if $i>r$.

3. $a^r_i=0$ if $r$ and $i$ have  different parity.

4. Consider $a^r_i$ as a function of the variable $n$. Then
$a^r_i(n)$ is a polynomial of degree $(r-i)/2$ as $r$ and $i$ have
the same parity. Moreover, $a^r_i(n)=
C(r,r-i)n^{(r-i)/2}+p_{r,i}(n)$ where
$C(r,s)=\frac{(r+s)!(r-s+1)}{s!(r+1)!}$ are entries of Catalan's
triangle and $p_{r,i}$ is some polynomial of the degree  at most
$\frac{r-i}{2} -1$.

 Notice
 that $a^r_i$ is the number of the paths of the length $r$ in the
hypercube such that distance between the origin and the end of the
path is equal to $i$. It is known that
$a^r_i=(\cosh^{n-i}(x)\sinh^{i}(x))^{(r)}|_{x=0}$ and

$a^r_i=\frac{1}{2^n}\sum\limits_{k=0}^n(n-2k)^rP_k(i,n)$ where

$P_k(r,n)=\sum\limits_{j=0}^k(-1)^j{{r}\choose{j}}{{n-r}\choose{k-j}}$
are the $k$th Krawtchouk polynomials of the variable $r$.

Define $\nu_k(n)=\max \nu_k(S)$ where  $S\subset F^n_2$,
$|S|=2^{n-k}$.

\begin{theorem}
 Let $r\leq k$ be even. Then $\nu_k(n)\leq 1-
\frac{1+o(1)}{2^{k+1}}{{k}\choose{r}}$ as $n\rightarrow\infty$.
\end{theorem}

Proof. Let $r\leq k$ be even. Consider  $S\subset F^n_2$,
$|S|=2^{n-k}$, and let $f$ be the characteristic function of $S$.
For any function $g\in\mathbb{V}$ it holds

$(M^rg,g)=\sum\limits_{t=0}^n (n-2t)^r
\sum\limits_{wt(z)=t}(\phi_z,g)^2$.

Since $(\phi_{\overline{0}},f)=|S|/2^n= 2^{-k}$, we  obtain
$(M^rf,f)\geq n^r2^{-2k}$.

It is easy to see that $M_i\phi_{\overline{0}}={n \choose
i}\phi_{\overline{0}}$. Then it holds $(M_if,f)\leq {n \choose
i}(f,f)$ . Hence $(a_i^rM_if,f)\leq a_i^r{n \choose
i}=O(n^{i+(r-i)/2})$ as $n\rightarrow\infty$.

Since $a^r_r=r!$, we obtain

\begin{equation}\label{evid2}
(M_rf,f)=\frac{n^r(1+o(1))}{r!2^{2k}}.
\end{equation}

By definition of $M_r$ the number of the pairs from $S$ at the
distance $r$ is equal $2^{n-1}(M_rf,f)$. Since a pair of points at
the distance $r$ lies in ${{n-r}\choose{k-r}}$ $k$-faces, the number
of the $k$-faces without points from $S$ is not less then
$2^{n-1}(M_rf,f){{n-r}\choose{k-r}}$. Then $1-\nu_k(S)\geq
\frac{2^{n-1}(M_rf,f){{n-r}\choose{k-r}}}{2^{n-k} {{n}\choose{k}}}$
where ${2^{n-k} {{n}\choose{k}}}$ is the number of the $k$-faces.
Thus, the statement follows from (\ref{evid2}). $\triangle$

Let $r(k)$ be the nearest even number for $k$. Then
$\frac{1}{2^{k+1}}{{k}\choose{r(k)}}=\frac{1+o(1)}{\sqrt{2\pi k}}$
as $k\rightarrow\infty$. For example,
$\lim\limits_{n\rightarrow\infty}\nu_3(n)\leq \frac{13}{16}$.

\section{Linear codes}

The results of this section were announced in \cite{FRP}.

\begin{pro} Let $C$ be a linear code and $\Gamma$ be a $k$-face.
Then it is possible two cases

1) $|C\cap (x+\Gamma)|=0\ {\rm or}\ 2^s$ for each $x\in F_2^n$;

2) $|C\cap (x+\Gamma)|=1$ for each $x\in F_2^n$.

\end{pro}

Proof. System of linear equations over $GF(2)$ has $1$ or $0$ and
$2^s$ solutions for different right-hand side vectors.
 $\triangle$

Denote by $\mu_k(S)$ the number of $k$-faces containing  only one
element of a linear code $S$ (this element is $\overline{0}$). It is
clear that $\nu_k(S) > \mu_k(S)/{{n}\choose{k}}$.

Let $H=\{h_{ij}\}$ be a binary $k\times n$ matrix. Consider the
linear code $C=\{x\in F_2^n | Hx=\overline{0}\}$. The characteristic
function $\chi^C$ of $C$ is equal to $\prod_i(1\oplus \bigoplus_j
h_{ij}x_j)$.

 \begin{pro}\label{pvc2} The number of monomials of the degree $k$ in
$\chi^C$ is equal to $\mu_k(C)$.  \end{pro}

It follows from the properties of M\"obius transform of Boolean
function.

Let $H[x]$ be a matrix with columns $h_jx_j$ where $h_j$ is a column
of $H$, $j=1,\dots,n$. For example, if $ H_2= \left(
 \begin{array}{ccc}
0 & 1 & 1\\
1  & 0 & 1\\
  \end{array} \right)$ then $ H_2[x]= \left(
 \begin{array}{ccc}
0 & x_2 & x_3\\
x_1  & 0 & x_3\\
  \end{array} \right)$.

  The permanent of the matrix $H$ (square or rectangular $k\times n$, $k\leq n$) is the sum of
  all products $h_{1j_{1}}\dots h_{kj_{k}}$ where $j_1,\dots,j_k$
  are mutually different. There we use addition by modulo 2. For example,
  ${\rm per}_2(H_2[x])=x_1x_2+x_2x_3+ x_1x_3$.

  By the definition we can obtain the following properties of ${\rm
  per}_2H[x]$.

  1. If we add a row of $H[x]$ to another row of $H[x]$ then ${\rm
  per}_2(H[x])$ does not change.

  2. Laplace expansion for determinants is true for the permanents of rectangular
  matrices
  (decompositions by row only).

  3. If two columns $h_{n-1}$ and $h_n$ of $H$ coincide, then ${\rm
  per}_2(H[x_1,\dots, x_n])$ is equal to

  ${\rm
  per}_2(H'[x_1,\dots, x_{n-2}, x_{n-1}+x_n])$ where $H'$ is  $H$
  without the last column.

These properties are proved analogously to the property of the
determinant
  because the permanent and the determinant coincide  modulo 2.

Let $H_k$ be the matrix $k\times (2^k-1)$ such that the $j$th row of
$H_k$ is the binary representation $b_k(j)$ of  $j$. Determine real
functions $f_k$ by the equations $f_k(x_1,\dots, x_n)={\rm
  per}_2(H_k[x_1,\dots, x_n])$, where $n=2^k-1$.

\begin{pro}
$f_k(x_1,\dots, x_n)=\sum\limits_{i_1,\dots,i_k\in I} x_{i_1}\cdots
x_{i_k}$,\\ where $I= \{i_1,\dots,i_k \}$ and vectors $b_k(i_1),
\dots, b_k(i_k)$ are linearly independent over $GF(2)$.
\end{pro}

Let $H$ be a  binary $k\times n$ matrix having $a_j$ columns
$b_k(j)$ for $j=1\dots 2^k-1$. Consider the linear code $C=\{x\in
F_2^n | Hx=\overline{0}\}$. Applying properties of permanents we can
prove

\begin{pro} $\mu_k(C)=f_k(a_1,\dots,a_{2^k-1})$.
\end{pro}

Proof. By Proposition \ref{pvc2} and the third property of the
permanent
$$\mu_k(C)={\rm
  per}_2(H[x_1,\dots, x_n]|_{x=\overline{1}}={\rm
  per}_2(H_k[\underbrace{x_{i_1}+\dots}_{a_1},\dots,
  \underbrace{x_{i_n}+\dots}_{a_{2^k-1}}]|_{x=\overline{1}}=f_k(a_1,\dots,a_{2^k-1}).$$

Then we can conclude that the original problem is equivalent to
finding maximum of  $f_k$ where the sum of the arguments is fixed.
Consider the set
$$T_k(s)=\{(x_1,\dots,x_n) | x_j\geq 0,
\sum\limits_{j=1}^nx_j=s\},$$ where $n=2^{k}-1$.

  \begin{theorem} $\max\limits_{x\in
  T_k(s)}f_k(x)=f_k(s/n,\dots,s/n)$.  \end{theorem}

  Proof by induction on $k$. For $k=2$ the statement of the theorem can be
  verified by the standard methods of analysis.
Consider occurrences of the variable $x_1$ in the polynomial
$f_k(x_1,\dots, x_n)$. By the second  and the third properties of
the permanent we get

$${\rm
  per}_2(H_{k+1}[x_1,\dots, x_{2n+1}])= x_1{\rm
  per}_2(H_k[(x_{i^2_1}+x_{i^3_1},
  \dots, x_{i^{2n}_1}+x_{i^{2n+1}_1})]) + h(x_2,\dots,x_{2n+1}), $$ where a function $h$ does
  not depend on the variable $x_1$,  $b_k(i^{2j}_1)$ and
  $b_k(i^{2j+1}_1)$  differ only in the first position.

 By the first property of the permanent all variables of $f_k(x_1,\dots,
 x_n)$ are equivalent and each monomial has degree
 $k$.
   Hence we have the equation

  $$f_{k+1}(x_1,\dots,x_{2n+1}) =
  \frac{1}{k+1}\sum\limits_{j=1}^{2n+1}x_jf_k(x_{i^1_j}+x_{i^2_j},
  \dots, x_{i^{2n-1}_j}+x_{i^{2n}_j}),$$ where  $\{x_{i^m_j}\}$ is
  the set of all variables without $j$.

 By induction, we find $f_{k+1}(x_1,\dots,x_{2n+1})\leq
\frac{1}{k+1}\sum\limits_{j=1}^{2n+1}x_jf_k((s-x_j)/n,
  \dots (s-x_j)/n).$ Determine $y_j=s-x_j$.
  Obviously we have $\sum\limits_{j=1}^{2n+1}y_j=2ns$. Then we obtain
 an inequality
$f_{k+1}(x_1,\dots,x_{2n+1})\leq
\frac{f_k(\overline{1})}{k+1}\sum\limits_{j=1}^{2n+1}
(s-y_j)\left(\frac{y_j}{n}\right)^k$.

By the method of Lagrange multipliers, it is not complicated to
prove that the function $g(y_1,\dots,y_{2n+1})=
\sum\limits_{j=1}^{2n+1} (s-y_j)y_j^k$ has maximum in the interior
point $y_1=\dots=y_{2n+1}=\frac{2ns}{2n+1}$ if
$\sum\limits_{j=1}^{2n+1}y_j=2ns$, $0\leq y_j\leq s$. In the edge
points (where $y_i=0$ for some $i$, $i=1,\dots, 2n+1$), this
function isn't positive. Induction step is proved. $\triangle$

\begin{corol} For fixed $k$ and $n=m(2^k-1)$ the maximum value of
$\mu_k(C_{k,m})$ corresponds to the code  $C_{k,m}$ with  the check
matrix $H$ consisting of $m$ columns $b_k(j)$ for all $j=1,\dots,
2^k-1$. If $m=1$ then $C_{k,1}$ is the Hamming code.\end{corol}

For example, $H=(1 1 1 \dots 1)$; $H= \left(
 \begin{array}{ccc}
0 & 1 & 1\\
1  & 0 & 1\\
  \end{array} \right)$; $H=\left(
 \begin{array}{cccccc}
0 & 0 & 1 & 1& 1 & 1\\
1 & 1 & 0 & 0 & 1 & 1\\
  \end{array} \right)$;

$H=\left(
 \begin{array}{ccccccc}
0 & 0 & 1 & 1& 1 & 0 & 1\\
0 & 1 & 0 & 0 & 1 & 1 & 1\\
1 & 0 & 0 & 1 & 0 & 1 & 1\\
  \end{array} \right)$.

By linear algebra it is routine to prove the following proposition.

\begin{pro} Let $n=2^k-1$, $k\geq 2$ then
$\mu_k(C_{k,1})=(2^k-1)(2^k-2)(2^k-4)\cdots(2^k-2^{k-1})/k!$ and

$\nu_k(C_{k1})=\frac{1}{k!2^k {n \choose
k}}\sum\limits_{t=1}^{k-1}(2^k-1)(2^k-2)(2^k-4)\cdots(2^k-2^{t})
{{2^{t+1}-t-2}\choose{k-t-1}}2^{t+1}$.

\end{pro}

It is possible to calculate that $\nu_2(C_{2,1})=1$,
$\nu_3(C_{3,1})=9/10$, $\nu_4(C_{4,1})=10/13$.


 Define $\nu'_k(C)=\mu_k(C)/{n\choose
k}$. Obviously, $\nu'_k(C)$  is a lower bound of $\nu_k(C)$. Then
$\nu'_k(C_{k,1})
=(2^k-1)(2^k-2)(2^k-4)\cdots(2^k-2^{k-1})/(2^k-1)(2^k-2)(2^k-3)\cdots(2^k-k)$
 and
$\lim\limits_{k\rightarrow\infty}\nu'_k(C_{k,1})\approx 0.2888$.

\begin{corol}
$\lim\limits_{s\rightarrow\infty}\nu'_k(C_{k,s})=(2^k-1)(2^k-2)(2^k-4)\cdots(2^k-2^{k-1})/(2^{k}-1)^k$.
\end{corol}

\section{Conclusion}

As we can see from Proposition \ref{pvid10} the best (for our task)
linear code is not better than a random set as $k$ is large. But if
$k$ is a small integer than the best linear code is tight to the
best unrestricted partial covering array. Problems to find the best
unrestricted set or to find an asymptotic of its cardinality are
open.

\end{document}